\relax
\documentclass{article} 
\usepackage[square,sort,comma,numbers]{natbib}
\usepackage{arxiv}  
\usepackage{times}  
\usepackage{helvet}  
\usepackage{courier}  
\usepackage{url}  
\usepackage{graphicx}  
\usepackage{url}
\usepackage{graphicx,epsfig}
\usepackage{subfig}
\usepackage{graphics,dcolumn,bm,fleqn,float}
\usepackage{amssymb,amsmath,multirow,rotate,color}
\usepackage{epstopdf}
\usepackage{soul}
\usepackage{listings}
\usepackage{xcolor}
\usepackage{booktabs}
\usepackage{natbib}
\usepackage[T1]{fontenc}
\usepackage{xspace}
\usepackage[flushleft]{threeparttable}
\usepackage{booktabs}
\usepackage{balance}
\usepackage[inline]{enumitem} 

\newlist{RQ}{enumerate}{1}
\setlist[RQ]{label=\textbf{RQ}-\arabic*,resume,leftmargin=*}

\renewcommand{\footnotesize}{\scriptsize}

\newcommand\reftab[1]{Table~\ref{#1}}

\def\ie{\emph{i.e.,}\xspace}

\newcommand{\incoming}{$\Leftarrow$}
\newcommand{\outgoing}{$\Rightarrow$}

\begin{document}
%
\title{Tweeting MPs: Digital Engagement between Citizens and Members of Parliament in the UK}

\author{
 	Pushkal Agarwal$^1$,  Nishanth Sastry$^1$,  Edward Wood$^2$ \\ 
     $^1$King's College London, $^2$House of Commons Library\\
     $^{1}$\{pushkal.agarwal,nishanth.sastry\}@kcl.ac.uk, $^2$woode@parliament.uk\\
}

\maketitle

\begin{abstract}

Disengagement and disenchantment with the Parliamentary process is an important concern in today's Western democracies. Members of Parliament (MPs) in the UK are therefore seeking new ways to engage with citizens, including being on digital platforms such as Twitter. In recent years, nearly all (579 out of 650) MPs have created Twitter accounts, and have amassed huge followings comparable to a sizable fraction of the country's population. This paper seeks to shed light on this phenomenon by examining the volume and nature of the interaction between MPs and citizens. We find that although there is an information overload on MPs, attention on individual MPs is focused during small time windows when something topical may be happening relating to them. MPs manage their interaction strategically, replying selectively to UK-based citizens and thereby serving in their role as elected representatives, and using retweets to spread their party's message. Most promisingly, we find that Twitter opens up new avenues with substantial volumes of cross-party interaction, between MPs of one party and citizens who support (follow) MPs of other parties. 


\end{abstract}

\section{Introduction}
\label{sec:intro}



There has been much bemoaning the apparent decrease in political engagement amongst the electorate in western democracies such as the UK~\citep{parliament}.  Connecting elected representatives such as Members of Parliament (MPs) with voters (especially young voters) is seen as a means to ``revive democracy''~\citep{cookMP} and recently there is much hope that online methods such as Twitter will play a key role in this~\citep{openup}. Yet, traditional scholarship on legislative studies has focused mostly on the relationship between the Parliament and the Government, casting MPs in the core roles of legislation and scrutiny of the Executive branch, neglecting the communication between citizens and their MPs~\citep{leston2012studying}. 

In this paper, we are interested in characterising how engagement of democratic representatives with their citizens is shaped by online platforms, more specifically Twitter. We focus on the UK,  where a remarkable 579 out of 650 Members of Parliament (MPs) are active on Twitter. This represents a dramatic rise from just a few years back: only in 2011, we had just 51 MPs ``dipping their toes'' in Twitter~\citep{jackson2011microblogging}. Now, however, these MPs, who represent a nation of 65 Million, have a collective following of 12.83 Million on Twitter (some of these users follow multiple MPs; the total number of unique users following at least one MP is 4.28 Million.). Thus, Twitter appears to have become a platform on which MPs can engage with a substantial number of citizens. 

To frame the discussion, we consider the nearest offline equivalent for interaction between MPs and citizens -- \emph{constituency service}. The traditional means by which this is done is for the elected representatives to hold open and private meetings with those that elected them. In the UK, for example, MPs travel back to their constituencies, typically on Thursdays, after the work of the Parliament is done, and hold `surgeries' with their constituents. `Town hall' meetings in the USA serve a similar purpose.  Constituents may also phone or email their MPs and members of Congress to let them know their positions on key issues.  

To be sure, there are differences between engagement on Twitter and traditional constituency service. Twitter interaction can be immediate and spontaneous, in contrast with scheduled surgeries. The public nature of Twitter renders it unsuitable for constituency service requiring personal information. In our dataset in 1548 cases, MPs asked to move away from public discussions on Twitter, asking constituents to make appointments at their surgeries, offering their email addresses or asking the respondent to ``DM'' or ``direct message'' them. In a handful ($\approx 10$) of cases, both options were offered\footnote{One MP wrote: ``@XX, If you follow me I'll DM you or please email YY@ZZ and I will get back to you. Thanks, D''}. These interactions represent over 7\% of replies by MPs. Also, constituency service is usually seen as MPs engaging with and serving those in the \emph{geographic area they represent}. In the UK, there is even a strict parliamentary protocol that MPs do not seek to intervene or act in matters raised by the  constituents of other members~\citep{protocolDPA}. On Twitter, however, it can be hard for MPs to tell the precise location of their correspondents, and the immediacy and public nature of the medium may lead to interactions with non-constituents. 

Despite such differences, both forms of communications hold the same promise: direct contact and engagement between elected officials and those they are supposed to represent. Therefore, we turn to the literature on constituency service interactions for pointers on the nature of the discourse between citizens and UK MPs on Twitter. The traditional view of psephologists has been that constituency service is worth only about 500 votes~\citep{norton1990constituency,butler2001payment}, and thus, is insufficient to make a difference in all but the closest of elections. \cite{king1991constituency,krasno1997challengers} talk about the `incumbency factor' and the need for MPs to develop this relationship in order to get re-elected. Thus, more than being a campaigning tool, engaging with citizens can be seen as a mechanism for building relationships and achieving better representation. 

Based on these considerations, we focus on a two-month period from Oct 1, 2017 to Nov 29, 2017, when there was no election going on, and thereby seek to understand the usage of Twitter as a tool for everyday citizen engagement. The period also encompasses times when Parliament was in session (requiring MPs to be away from their constituencies, attending the House of Commons) and in recess\footnote{https://www.parliament.uk/about/faqs/house-of-commons-faqs/business-faq-page/recess-dates/} (when MPs are free to return to their constituencies), and therefore can be expected to cover both aspects of MP activities. We study the Tweets, Retweets and Replies of MPs towards other users, as well as from other users towards MPs. Both groups are active, with the MPs and the citizens respectively producing 178,121 and 2,339,898 Tweets, Retweets and Replies directed at each other.

The parallels and distinctions between engagement on Twitter and constituency work also drive our research questions: Norton and Wood's seminal study of British MPs' constituency work in the 80s concluded that constituency service can  be extremely rewarding, although taxing, taking MPs close to ``saturation point''~\citep{norton1993back}. Therefore it is natural to ask whether Twitter imposes a burden on MPs. Given that any additional work would also likely take time away from other duties of the MP, we also wish to understand how MPs manage whatever burden is imposed on them by their Twitter presence. Following the typology of \cite{stanyer2008elected} who studied how MPs use constituency service to package themselves, we ask whether MPs are using Twitter to prioritise helping constituency members, gaining personal visibility by highlighting work they have done, for spreading the message of their party and party leaders, or for other purposes. Finally, we are interested in identifying the tone of the conversation online. Given the tendency of Twitter as a  polarising and sometimes aggressive sphere \citep{chatzakou2017mean,conover2011political}, we ask what the nature and tone of the conversation is, between MPs and others. This discussion can be crystallised into the following research questions:
\begin{RQ}
\item  As a new and additional medium of citizen engagement, how much load does Twitter place on MPs, and how does this load vary?
\item How do MPs manage the load imposed? Do they selectively prioritise certain forms of engagement or seek external help (e.g., from their staff)?
\item What is the nature and tone of the conversations? Is Twitter a polarising sphere with echo chambers for each party and side of the political spectrum? Is the tone civil or aggressive?
\end{RQ}

We find that attention to individual MPs varies dynamically: although there is a huge amount of information overload during short periods of time which we term as ``focus windows'', there is a significant amount of ``churn'' in the set of MPs who are ``in focus'' at any given time. MPs strategically manage their relationship with their followers and this information overload by balancing their different roles as representatives of their constituency and their party~\citep{stanyer2008elected}: They selectively reply to Twitter profiles in the UK and within their constituency region, fulfilling their representative role, and use retweets as a mechanism to promote their image and spread the message of their party. Interestingly, we find evidence of significant cross-party interaction, between citizens who support and follow MPs from one party, and MPs from other parties. Thus, in an atmosphere of growing political divide in the UK (e.g., \citep{times,wpost}), Twitter seems to offer ways to avoid the ``echo chamber'' behaviour which characterises much consumption of information about politics online.

\subsection{Related Work}

Earlier studies on the use of Twitter by UK politicians mostly relate to a period when such usage was in its infancy, with a small fraction of MPs being regular users~\citep{jackson2011microblogging, lilleker2013online, graham2013between, graham2016new}.  Nevertheless, there were early indications that use of Twitter was entering the mainstream of electoral campaigning and political communications generally. Although, political tweets might not look like substantive \emph{contributions} to the political discourse, they appear to have become \emph{an increasingly integrated element of political communication in a `hybrid media system'}~\citep{jungherr2016twitter}.

In earlier usage of Twitter, one-way communications (``broadcasting'') predominated  ~\citep{graham2013between,jackson2011microblogging,lilleker2013online}, but participatory communication through Twitter was seen to be emergent.  It seemed to fit neatly into ~\cite{coleman2005blogs}'s concept of direct representation ~\citep{graham2013between} and politicians talked about their use of Twitter in these terms, but it was still secondary to other uses~\citep{jungherr2016twitter}.  Nevertheless, \cite{graham2013between} found that 19\% of candidates' tweets during the 2010 general election campaign interacted in one way or another with voters, which they argued was a fairly substantial level of interaction compared to other forms of political communication during the campaign.  A participatory style of communications on Twitter had potential to earn legislators political capital ~\citep{jackson2011microblogging} and was the only statistically significant strategy that had a positive impact on the size of the community ~\citep{lilleker2013online}.

Cross-party communication \emph{between MPs} was found to be unusual.  Unsurprisingly, MPs indulged in one-off attacks on other politicians during the 2010 UK general election campaign~\citep{graham2013between} but there was evidence of a more collaborative approach amongst an ``organic community'' of early adopters on Twitter~\citep{jackson2011microblogging}. Supporters of different parties tended to cluster around different hashtags during election campaigns, creating politically separated communication spaces ~\citep{jungherr2016twitter}. Here, we focus on communications \emph{between MPs and non-MPs} during periods when there is no election, and find that cross-party talk is more prevalent.

Our focus on a period without an election also makes our efforts complementary to the large number of works that examine the (ab)use of Twitter during election campaigns~\citep{graham2013between,anstead2015social,graham2016new,jungherr2016twitter,allcott2017social,koc2016normalization,lilleker2017drives}. Complementary to our focus on MP efforts, \cite{lilleker2017drives} studies  citizens' participation in political discussions online, and finds that extrinsic motivations such as social norms are the most significant in mobilising efforts.

\section{Background and dataset} \label{sec:dataset}
In this section, we give a background about the UK democratic process, focusing on MPs and their communication needs and motivations, and describe the datasets\footnote{The dataset we collected is made available at https://nms.kcl.ac.uk/netsys/datasets/tweeting-mps/ for non-commercial research usage. Following Twitter's Terms of Usage  (\url{https://twitter.com/en/tos}), we will only be able to share the Tweet IDs.} we have collected to answer our questions.   

\subsection{MPs and Democracy in the UK}

The Parliament in Westminster is the supreme legislature in the UK. It is composed of two houses or chambers. The primary house is the House of Commons. It has 650 elected members. Most MPs at any given election are drawn from a handful of major political parties. It is possible for candidates to run for election without the backing of a political party but they are very unlikely to get elected. The three major parties in the UK Parliament following the 2017 general election were Conservative (Cons.), Labour (Lab.) and Scottish National Party (SNP). Other  parties with MPs include the Liberal Democrats (Lib Dems) and the Democratic Unionist Party (DUP).
 
 Members of Parliament are elected on a ``party ticket'' or manifesto and when they vote in the House of Commons they are expected to obey party discipline.  This also applies to their publicity and engagement work, where they are discouraged from giving messages that are inconsistent with the party line. This role of the MP as a party representative may sometimes conflict with the role of MPs as representatives of their constituencies. However, some MPs are more loyal to their party than others~\citep{cowley2002revolts}, and in some cases, may choose constituency over party.

\subsection{Datasets}
\paragraph{MPs on Twitter} 
We start with the MPs who are active on Twitter, as obtained from a comprehensive and up to date list\footnote{http://www.mpsontwitter.co.uk/list}. The UK House of Commons has 650 Members of Parliament (MPs). Of these, 579\footnote{we have collected data for 559 MPs since last year, and have not included the 20 new MPs who have joined since then.} MPs (187, i.e., 32.37\% are female) are active on Twitter, with a total of $\approx$ 13 Million followers. 
For each MP, we obtained the following data: 
\begin{description}
\item[Follower and Following] Using twitter API, we fetched all the users ($\approx$ 4.28 Million) who follow MPs and also the users that MPs followed (869K).
\item[Tweets and replies (\outgoing)] Using the Twitter API, we obtain from the MPs' Twitter timelines a total of up to 3,200 original tweets, retweets and replies to other Twitter handles.This covers the period of Oct 1 - Nov 30 2017, and we are able to fetch all MPs' timelines within the maximum limit of 3,200 allowed by Twitter. 
These collectively identify the utterances made \textit{by} the MPs, directed \textit{towards other} Twitter users. We will use the symbol $\Rightarrow$ to refer to such Tweets. 
\item[Mentions and replies to MPs (\incoming)] 
To fully understand the extent of the conversation, we obtain the utterances of all \textit{other} Twitter users\footnote{In the rest of this paper, we interchangeably use the terms ``ordinary'' Twitter users and ``citizen'' to refer users who have mentioned an MP in one of their Tweets. When the term citizen is used, it has been verified (if relevant), that the users included are those who declare a profile location in the UK (See Geography details above).}, directed \textit{towards} the MPs. This is obtained by searching for the MPs' Twitter handles using Twitter's  ``advanced search'' API, and includes all mentions of the MPs' Twitter handles, whether as a reply to a tweet of an MP, or merely mentioning an MP's Twitter user name in a non-reply Tweet. We will use the symbol $\Leftarrow$ to refer to such Tweets. 
\vspace{-0.5cm}
\end{description}
\begin{center}
\begin{table}
	\centering
	\begin{tabular}{ |p{3.5cm}|p{4cm}| } 
	\hline
	MPs on Twitter     & 559    \\ 
	\hline
	Verified MPs    & 83.54\%  \\
	\hline
	Total Followers   & 12.83 Million  \\
	\hline
	Mentions per MP per day ($\Leftarrow$) & Mean: 67.96, \newline Median: 13.25, \newline Standard Deviation: 302.82 \\
	\hline
	Activity (Tweets, replies to others) per MP per day ($\Rightarrow$) & Mean: 5.52, \newline Median: 3.3, \newline  Standard Deviation: 6.18 \\
	\hline
	\end{tabular}
\caption{Details of the Twitter dataset (Oct 1 -- Nov 30 2017).}
	\label{tbl:dataset}
	      \vspace{-0.5cm}
\end{table}
\end{center}
Collectively,  $\Rightarrow$  and $\Leftarrow$  capture both sides of the conversation between MPs with Twitter handles and the rest of Twitter. To understand how people talk about MPs \textit{who are not on Twitter}, we searched for the full (first and last) names of such MPs, obtaining 35,904 Tweets. To ensure that this refers to MPs and not some other person with the same name, we manually examined all Tweets and compiled a list of the most politically related words used in conjunction with these names (\texttt{mp, Brexit, Parliament, Westminster, Tory, Minister, Party, Conservative, Labour, Vote, Democracy}). Filtering for these keywords, we  are able to retain 15,083 (of the  35,904) Tweets. Since these are Tweets mentioning MP names, we term these as ``\textit{pseudo-mentions}''. Clearly we are conservative in capturing pseudo mentions, and may have ignored several tweets, e.g., those that may not use the first and last names of the MPs. However,  pseudo-mentions introduce data about 67 MPs. Thus, information about \emph{646 of the 650 MPs are captured in our dataset, in one way or another.} Statistics about the dataset are listed in Table~\ref{tbl:dataset}.

We also perform additional heuristic processing to obtain the following information for MPs and all users who mention them or are mentioned by the MPs (through retweets, replies or original tweets). Where heuristics may lead to errors, we try to perform some checks, using limited ground truth to give some indication of the accuracy or coverage that we believe we have attained:
\begin{description}
	


\item[Geography] We assign country-level labels for each user as follows: Fetching the profile location of users, and checking for words such as UK, London, England, Wales, Scotland, Northern Ireland, United Kingdom, we labeled around half of the users. For the remaining ones, the  unique list of their locations is passed to Photon library's geocode function\footnote{http://photon.komoot.de/}, an open street API to label countries given location names (e.g., city/locality/state etc.) that a user might have used. Using this procedure, the countries of $\approx$ 82.7\%  (77.3\%) of users retweeted (replied to) by MPs were obtained.
	
	For users in the UK, we dig deeper. A Twitter profile location may mention only a city name (such as `Cambridge, UK'), rather than a specific constituency (such as `Cambridge South', which contains a small part of the City of Cambridge and close-by villages). Also, citizens may work in one constituency and reside in another which is close by. Thus, to identify interactions between a user and their local MP, we translate user locations to the `postcode area' which represents the region (e.g., `CB' represents Cambridge in Cambridge-related postcodes), and consider all interactions between users from that postcode area, and the MPs representing that postcode as interactions between MP and a potential constituent. 


\item[Party affiliation] For every user who has mentioned an MP, we associate the party affiliation of the MP with the user. Users have mentioned a mean (median) of 3 (2) MPs, from an average (median) of 2 (1) parties. We affiliate each user with one party. For users who have mentioned MPs from more than one party, we assign them the party they have mentioned the most. \textit{Check:} By checking for the presence of party names in the profile description amongst a sample of $\approx$ 8K Conservative and Labour supporters, we are able to correctly label nearly 7,400, yielding a 91.4\% accuracy for our heuristic. 
\end{description}

\section{Dynamics of Citizen Attention} \label{sec:attention}
In this section, we approach the first research question, and estimate the burden caused to MPs by their Twitter presence, by studying tweets directed \emph{towards} MPs by other Twitter users. Our starting point is the stark difference in \reftab{tbl:dataset} between the  number of  mentions that an MP gets (marked as \incoming), and the average number of tweets and reply activities made by them (marked as \outgoing). This suggests that MPs could be overloaded, and are not able to respond to all tweets directed at them. 

To examine this, we introduce metrics that measure the spread of attention load, in terms of mentions of MPs. We study the distribution of attention across time for any individual MP, and across all MPs during any given time window. We find that in any given time window, a small number of MPs are `in focus', and receive a large number of mentions. However, as the news cycle moves on, other MPs' activities come into focus. We then illustrate this phenomenon using examples and discuss the implications. 

\subsection{Attention is focused during small time windows }
To understand how overloaded the MPs are with the number of mentions they receive, we first examine high activity periods. We define a period of \emph{high activity} as a continuous sequence of days when the daily activity is considered as `high'. 
 Formally, given an MP $i$ and a threshold average number of mentions $T_i$, we define a continuous sequence of days $R$ as a \textit{high activity window} for MP $i$ if it satisfies the property $$high\_activity_i(R): \sum_{d \in R} v_{id} > T_i |R|$$ where $v_{id}$ is the number of mentions obtained by MP $i$ on day $d$. 
In this paper, we set the threshold for a high activity individually for each MP. A day qualifies as a `high activity' day  for an MP if the number of mentions received by the MP that day is higher than the \textit{personal} average for that MP\footnote{Other threshold definitions were examined, but not reported here due to space.}. Note that even if MPs only receive a large number of tweets during a short time window, this will increase their personal average for the whole 2 month duration of our data set. We term the longest continuous run of days during which an MP $i$ has more than his or her personal average number of Tweets mentioning them -- as their \textit{focus window} $T_i^{\max}$. We can  compute the fraction   $F(T)$ of  MP $i$'s mentions that are obtained during a time window $T$ as
$$F(T) = \frac{1}{V_{i}}  \sum\limits_{d \in T} v_{id}.$$ Here, $V_i$ is the total volume of Tweets mentioning MP $i$ in our dataset,  and $v_{id}$ is the number of mentions obtained on day $d$ in the window $T$. We define the \textit{Focus} of MP $i$ as the fraction of mentions $F_i = F(T_i^{max})$ obtained during the focus window $T_i^{\max}$. In other words, Focus measures what  fraction of an MP's mentions  during the  whole 2 months period covered by our data set is concentrated during the small Focus Window, \ie the longest continuous sequence of days during which the MP receives a higher than average number of Tweets.

\begin{figure*}[tb]
\centering
\subfloat[Focus] {			        				\includegraphics[width=0.4\textwidth]			{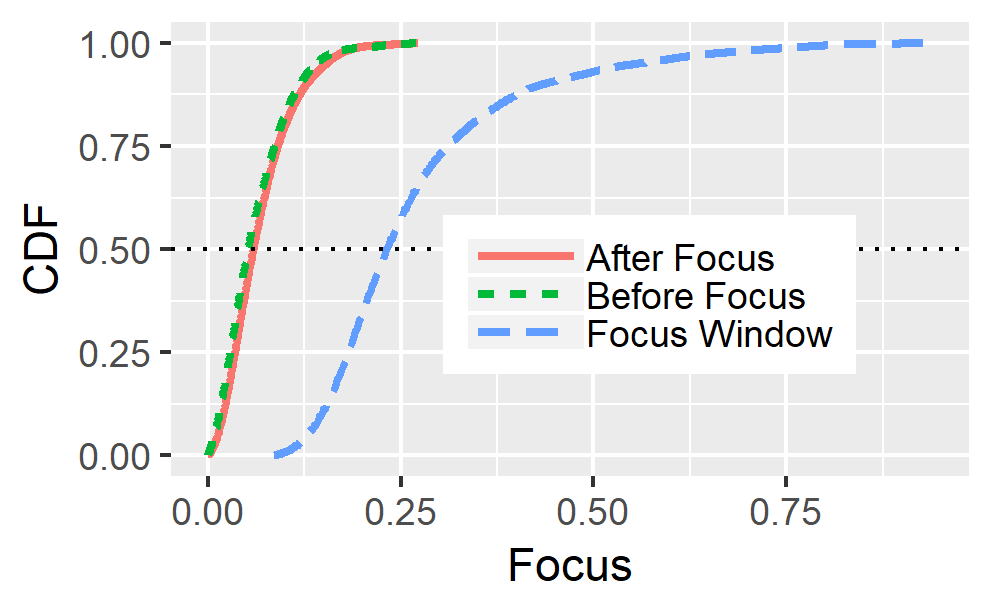}
	\label{fig:focus}}
\subfloat[Normalised Focus] {			        				\includegraphics[width=0.4\textwidth]			{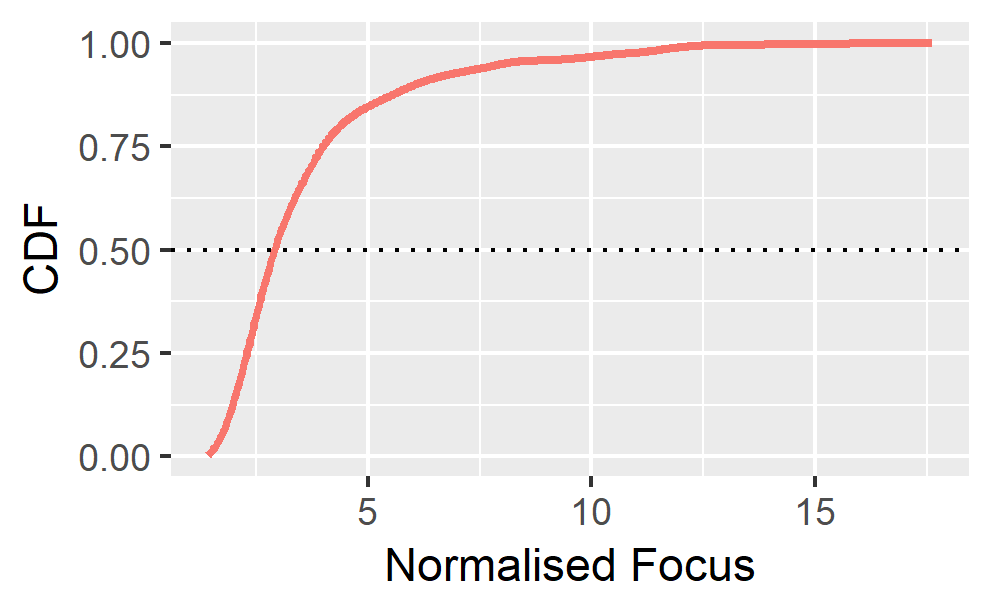}
	\label{fig:focusNorm}}\\
\subfloat[Churn]{
	\includegraphics[width=0.38\textwidth]		     {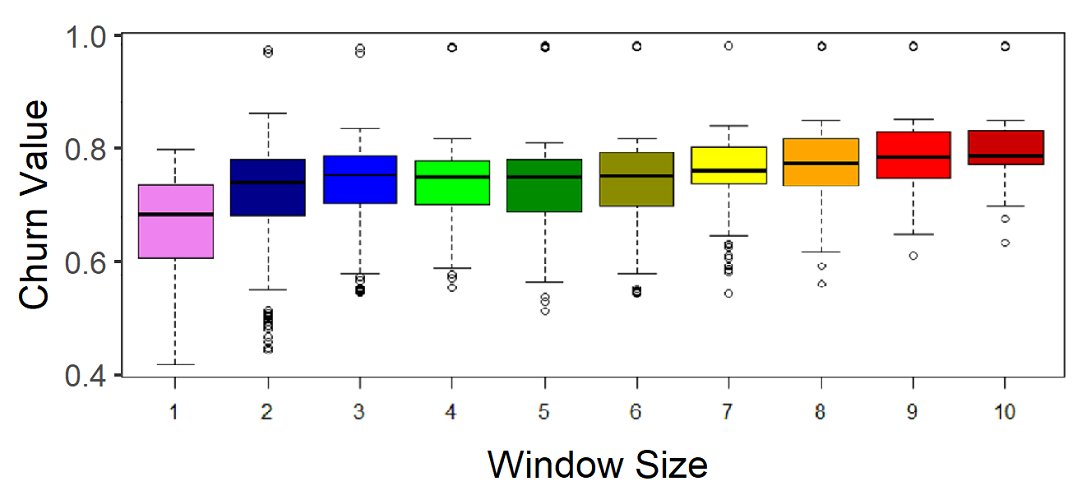}
    \label{fig:churn}}
\subfloat[Gini]{
	\includegraphics[width=0.36\textwidth]		 	{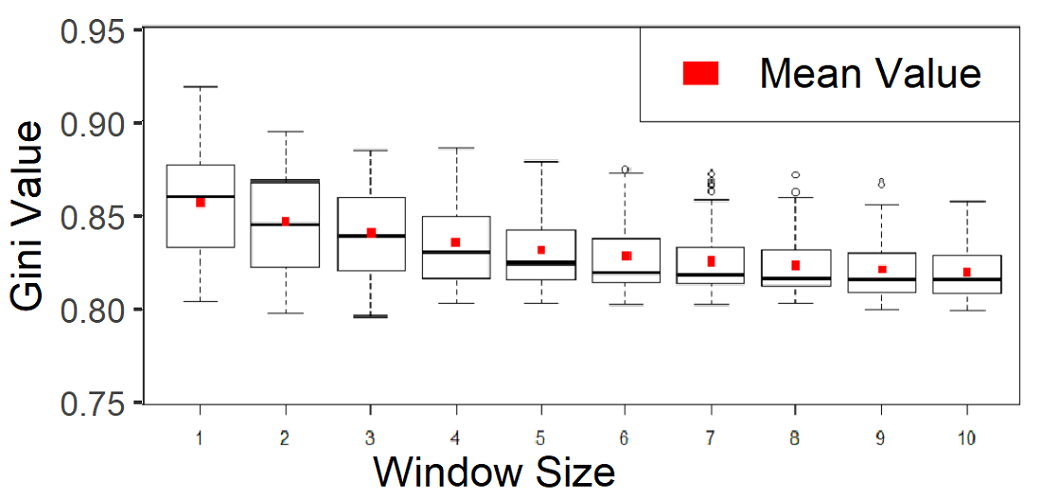}
	\label{fig:gini}}
\caption{The distribution of (a) \textbf{Focus} of MPs' mentions: cumulative distribution of  unnormalised fraction of mentions $F(T_i^{max})$ obtained by MPs during their focus windows. For comparison, the (much smaller) fractions of mentions $F(T_i^{b})$ and $F(T_i^{a})$ of  windows just before and after the focus windows is  shown.  (b) Cumulative distribution of Focus, normalised to yield a Focus of 1 if mentions were distributed evenly.  Most of the mass is several times over 1, confirming high information overload during focus windows.
(c) \textbf{Churn}: Box plots of the distribution of Churn values across  time windows of different sizes. The box extends from the lower to upper quartile values of the data, with a line at the median. The mean is shown as a green dot. Whiskers extend from the box to show the range. Flier points are outliers past the end of the whiskers. 
(d) \textbf{Gini}: Box plot of distribution of Gini co-efficients of the number of mentions received by MPs during  time windows of different sizes.  
}
 \vspace{-0.5 cm}

\end{figure*}

Figure~\ref{fig:focus} shows the distribution of Focus values for all MPs. For comparison, the fraction of mentions  $F(T_i^{b})$ and $F(T_i^{a})$ during similar-sized windows before and after the focus window is  also shown. Mentions tend to fall off rapidly outside focus windows: in the windows immediately preceding (following) these periods of intense activity, MPs on average receive less than  a quarter of the tweets received during the high activity focus window. 

By definition, Focus takes values between 0 and 1. We can get a sense of how  skewed focus values are by normalising the obtained focus based on the expected fraction of mentions given the size of the focus window: if the $V_i$ mentions are evenly across a total of $D$ days, the number of mentions expected in a focus window of $|R|$ days  is simply $|R|/D$. The observed Focus $F_i$ can therefore be normalised as $F_i D/|R|$. If mentions are uniformly distributed, normalised focus would be $\approx$ 1. Figure~\ref{fig:focusNorm} shows that this value tends to be several times larger than 1, suggesting that a disproportionately large fraction of the mentions for an MP might come during their one concentrated focus period. In other words, MPs are in the limelight only for a short period of time. Empirically, we find that the focus window period typically lasts between 3--5 days.

        


\subsection{Attention is unequal but focus moves among MPs}

The previous discussion suggests that  MPs receive mentions in a very bursty manner: Outside their focus window, an individual MP contributes much less to the overall volume of mentions directed towards MPs. Yet, as seen earlier, there is an average daily volume amounting to about 68 mentions per MP.  In this subsection, we look at how mentions are shared among the MPs on a daily basis. 

We proceed by considering all possible time windows of different sizes from 1--10 days. For instance, we can have 5-day windows from Oct 1--5, Oct 2--6 $\ldots$ Nov 24--29.  Our goal is to understand  the effect of MPs not receiving many mentions outside their focus windows and how mentions are shared during any given window.


For any time window of a given size, we ask how many of the high activity MPs of that window -- MPs who receive more than their personal average number of mentions --  continue to receive high numbers of mentions in the next window. 
 Formally, we define the set of active MPs during a time window $R$ as $$ active(R) = \{i | high\_activity_i(R)\}.$$  We can define the \textit{churn} of a time window $R$ and the time window $R^+$ immediately following it as the difference in the set of active users between the two windows: 
$$Churn(R) = \frac{|active(R) \triangle active(R+)|}{|active(R) \cup active(R+)|}$$ where the numerator is the symmetric set difference of MPs who are active in time window $R$ but not $R^+$ and vice versa, and the denominator is the union of users in the time windows. Figure~\ref{fig:churn} shows that  churn is high: Nearly 70\% of MPs who receive more than their personal average of mentions during one window are not able to sustain this  level of activity in the next window. Churn increases slightly as window sizes increase, with MPs finding it difficult to continuously receive high numbers of mentions over larger time windows.


While churn looks at differences across time windows, we can also measure how unequal the attention distribution is \emph{within} a time window, by counting the number of mentions each MP receives during the window and computing the gini co-efficient across all MPs receiving mentions. Gini co-efficients vary between 0.8 and 0.92, and the closer it is to 1, the more unequal the distribution being measured. We can get a sense of the inequality for different time windows of a given size by looking at the distribution of Gini co-efficients. Figure~\ref{fig:gini} shows the distributions of gini co-efficients for time windows of different sizes. The median gini co-efficient for all window sizes is consistently above 0.8, indicating that in any single window, most of the mentions are for a small minority of ``attention rich'' MPs.  

Collectively, these results suggest that during any given window, a few MPs are attention rich and receive a large number of mentions,  but this set of  MPs shifts over a period of days, so there are no overall ``superstars'' who are always at the centre of attention. The examples below serve to illustrate this phenomenon.

\subsection{Examples and implications}

\begin{figure}
	\centering
	\includegraphics[width=\columnwidth]{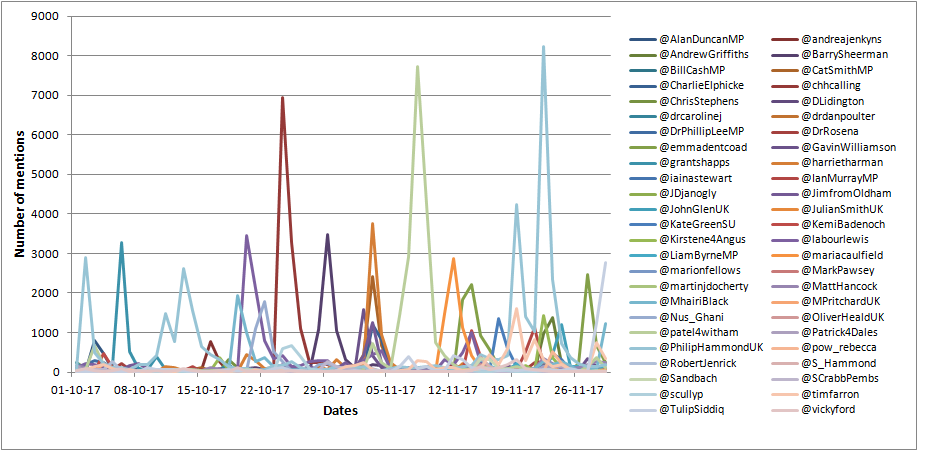}
	\caption{Timeline of number of mentions for 50 MPs who have more than half their mentions occur during their focus window.}
	\label{fig:highfocusMPs}
	 \vspace{-0.5 cm}
\end{figure}

To better visualise the attention imbalance, Fig.~\ref{fig:highfocusMPs} plots the daily mentions volumes of 50 MPs with the highest focus. The focus values for each of these MPs is more than 0.5; i.e., more than half of their mentions were received during their  focus windows. The spiky nature of the graph illustrates how attention can be highly concentrated during short focus windows (typically 3--5 days), and moves on to other MPs after the focus period.

Priti Patel (@patel4witham) represents an interesting example:  She was International Development Minister until 8 Nov 2017, but was forced to resign as a result of a scandal caused by unofficial meetings with Israeli ministers while on a holiday in that country. This resulted in a barrage of focused attention which fizzled out as other new stories cropped up. Similarly, Philip Hammond (@PhilipHammondUK), Chancellor of the Exchequer, had a huge number of mentions around the Autumn Budget (22 Nov 2017). Note that there is a smaller spike for Hammond just before the budget when he  mistakenly claimed in an interview that there are no unemployed in the UK\footnote{https://www.theguardian.com/politics/video/2017/nov/19/philip-hammond-there-are-no-unemployed-in-uk-andrew-marr-show-video}.  

These two examples illustrate two different kinds of focus windows: The attention towards Priti Patel was completely \emph{unanticipated} until the event unfolded, whereas the increased attention towards Philip Hammond as he presented the budget was predictable and could have been \emph{anticipated} and planned for (although even here, unanticipated mistakes can create spikes, as in Hammond's case). 

Anticipated attention is mostly for positive events and is in many cases  ``manufactured'' by the MPs, their staff and members of their party, following  prominent speeches or comments made in Parliament, as such successes are advertised by sharing widely on Twitter. A common source for such high attention events is activity during Prime Minster's Questions, which happens every single Wednesday at noon when the House of Commons is in session, and usually involves a lively and sometimes raucous debate. When an MP makes a particularly valuable (or sometimes particularly witty) contribution, it is shared by the MPs themselves, or by others, on Twitter, and then gets widely discussed.

By contrast, unanticipated focus windows, as with  Priti Patel, include mostly negative events for the MP. For instance, during the Westminster sex scandal, Charlie Elphicke (@CharlieElphicke), a Conservative MP, was accused of sexual misconduct and subsequently suspended from his party. The Westminster sex scandal, which coincided with the '\#MeToo' movement, includes several other resignations and castigations which also received high attention and focus values. Similarly, Labour MP Harriet Harman (@HarrietHarman) was criticised for mentioning an anti-semitic joke on live TV. 

The focus windows of 70\% (35/50) of the MPs  in Fig.~\ref{fig:highfocusMPs} are for events that could have been anticipated. 
However, perhaps unsurprisingly, unanticipated windows receive unusually high attention -- four of the top \emph{five} focus windows are apparently unanticipated, and for events which generated considerable adverse publicity. Thus, unanticipated attention can be all the more difficult to manage because of the volumes. Furthermore, all of the top five focus values are for MPs from the Conservative Party, which, as the current ruling party, tends to receive a large amount of scrutiny. Four of these also had ministerial level roles at one point or another and another held a senior role within the party. The fact that even such prominent MPs obtain more than half of their mentions during a small 3--5 day focus period illustrates that the attention of citizens is highly volatile and all too brief.

Focus periods represent opportunities for the MPs to raise their profile and engage with the populace on issues important to the MP. Whether the focus is a result of a positive event that the MP can take advantage of, or a negative event the MP should defend against, being able to appropriately handle the situation and manage the (brief) attention overload is critical. The next section looks at strategies that MPs use to manage citizens' attention both during their focus periods and out of their focus periods.

\section{Managing citizens' attention}

Incoming Tweets mentioning MPs (marked as \incoming\ in Table~\ref{tbl:dataset}) can be seen as a means for UK citizens and other Twitterati to engage with the MPs. In the previous section, we established that MPs faced an information overload with incoming tweets, especially during focus windows. In this section, we turn to the second research question, and ask how MPs manage this attention load in responding back, i.e., we also take into account the MPs' outgoing Tweets (marked \outgoing) in terms of Tweets, Retweets and Replies, and ask how MPs engage with the rest of Twitter. 

We identify two possible adaptations: The first consists of very selective replies, with MPs prioritising interactions with users local to their constituency region. The second is to employ staff who can help manage the load. We find extensive usage of the first strategy, with MPs largely prioritising their responses to users local to their region and to UK users. However, only some MPs appear to be using additional staff who can help manage their social media profiles.


\subsection{Selective replies and localism in MP actions}
\begin{table}
\centering
	\begin{tabular}{|p{2.7cm}|p{1.4cm}|p{1.4cm}|p{1.4cm}|} 
		\hline
		Geography & 
        \%Mentions $\Leftarrow$ &
        \%Retweet $\Rightarrow$&
        \%Reply $\Rightarrow$\\
		\hline
       UK \newline (Constituency (C)) &74.16 (C:28.8)&90.39 (C:56.7)&89.93 (C:59.04)\\
               \hline
        Commonwealth &3.88&2.35&1.74\\

        USA &5.34&3.42&4.63\\
        EU &3.47&2.53&1.72\\
        Others &13.15&1.29&2.06\\
        \hline
        Total &100\%&100\%&100\%\\
        \hline
	\end{tabular}
    \caption{Geographic Distribution of incoming mentions (\incoming) and outgoing actions (\outgoing) in different geographical regions. Each column adds up to a whole (i.e., UK + Commonwealth (British Commonwealth and Overseas Territories) + USA + EU + Others = 100\%). UK replies are further subdivided into replies within the constituency region of the MP, and those outside (The percentage local to MPs' constituency regions is shown in parenthesis as C:XX.YY\%). Thus, for instance, from (Row 1, Col 1), 74.16\% of all incoming (\incoming) mentions towards an MP come from within the UK. \textbf{Of these}, 28.8\% of mentions are from within each MP's constituency, and the remaining (71.2\%) are from outside their constituencies.}
    \label{selective-replies}
     \vspace{-0.5 cm}
\end{table}
MPs tend to be very busy, and being active online takes time away from their other duties, and their real-world constituency~\citep{jackson2008representation}. Therefore, we expect that MPs would be selective in who they respond to (even during non-focus periods). We test this hypothesis in two ways. First, we check the geographic areas of those whom the MPs respond to. Next, we check the category of the Twitter handles they respond to -- whether they are responding to other MPs, or those that they follow or are following them. 

Table~\ref{selective-replies} shows the percentage distribution of the incoming mentions (\incoming) and outgoing actions (\outgoing) -- replies and retweets, among different geographic regions.  In their conversations with Twitter users not from the UK, MPs tend to favour responding to Twitter users from countries that the UK has ties with:  USA (5.3\% mentions, 3.4\% retweets, 4.6\% replies),  British commonwealth and Overseas Territories (3.8\% mentions, 2.3\% retweets, 1.7\% replies) and the EU (3.4\% mentions, 2.5\% retweets, 1.7\% replies). Other countries get only 2\% of retweets or replies although they author over 13\% of tweets mentioning MPs.

As expected, a large fraction of mentions ($\approx$ 75\%) come from the UK, but MPs show selectivity, with over $\approx$ 90\% of their retweets and replies being made to UK-based Twitter users. This suggests that  \textit{Twitter is serving as a way for MPs to keep in touch with the UK electorate}. MPs are even more responsive to Tweets from within their constituency (identified as mentioned in the Dataset section). As shown in parenthesis in Table~\ref{selective-replies}, among incoming (\incoming) Tweets from within the UK that mention MPs, only about 28.8\% come from within the constituency region. However, MPs' outgoing (\outgoing) tweets prioritise interactions with such local tweets: 56.7\% of retweets, and nearly 60\% of replies are focussed within the constituency region represented by the MP\footnote{Note that this analysis only includes the 78\% of Tweets for which we are able to extract a valid geographic location of the Twitter profile with whom an MP is corresponding. We  also conservatively remove 60 London-based MPs from consideration because most MPs interact with journalists, lobbyists etc., who tend to be based around London. London MPs therefore appear to have an even higher localisation factor, with nearly all their responses to users within their region.}.

\begin{figure}
	\centering
	\includegraphics[width=0.5\columnwidth]{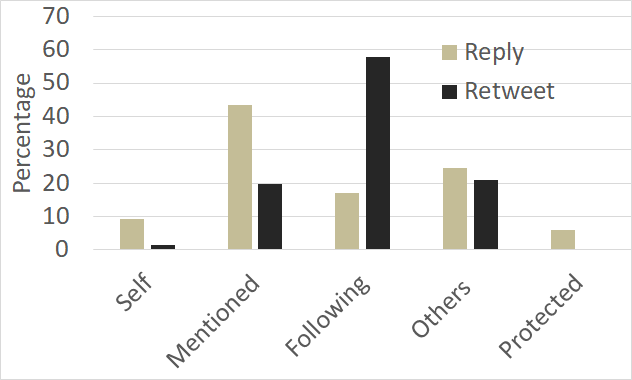}
	\caption{MP responses (retweets and replies) put into mutually exclusive categories. A `Self' response is a retweet or reply to MP's own tweet or reply. `Mentioned' is a response to a tweet mentioning the MP. `Following' is a response to a tweet which does not mention the MP but appears on their timeline because the MP follows the person. `Protected' tweets and replies are not available to analyse. `Others' comprises the remainder of retweets and replies.}
	\label{fig:ResponseType}
\end{figure}

We then compare the responses -- replies and retweets -- sent to different categories of people. Focusing first on the replies, Fig.~\ref{fig:ResponseType}  shows that $\approx$43\% of  replies are to tweets that mention the MP directly; thus MPs are using their replies to engage directly in conversation with those that mention them on Twitter. 
In contrast to replies, most (57.8\%) of the retweets  are for those that the MP follows. In other words, MPs are retweeting other users even without the MP being mentioned. This is not surprising, since Tweets from those that MPs follow appear on MPs' timeline, and  MPs may retweet what they find interesting. However, a disproportionate number of retweets are \emph{tweets of other MPs, and in particular, MPs from the same party}: on average, other MPs constitute 7.4\% of the following numbers of an MP. However, nearly 17\% of all retweet actions are made on Tweets of other MPs. A further 6\% of retweets are for posts made by their party's official Twitter handle. Nearly 96\% of the MP-MP retweets are for MPs  from the same party. Thus, it appears that \textit{MPs are using retweets as political marketing, to boost their party's message} (termed as party maintenance by \cite{stanyer2008elected}).

Figure.~\ref{fig:ResponseType} also shows that a small but significant minority of replies (9.4\%) are from the MP to themselves. This turns out to mostly be Tweetstorms -- a single post which has been split into series of related tweets (posted in quick succession) because of Twitter's character limit. On Nov 7 2017, close to the midpoint of our data collection period (Oct 1--Nov 29), Twitter did expand the character limit from 140 to 280, but this hardly affected the volume of self-replies: Prior to Nov 7, there was an average of 36.02 self-replies per day from all MPs, and after this date, the average was 34.17 per day. Thus, it appears that in many cases, MPs need a larger text limit than 280 characters to discuss substantive topics.

\subsection{Help from Staff?}

\begin{figure}[htbp] 
		\vspace{-0.5cm}
		\begin{center}
		\includegraphics[width=0.6\columnwidth]{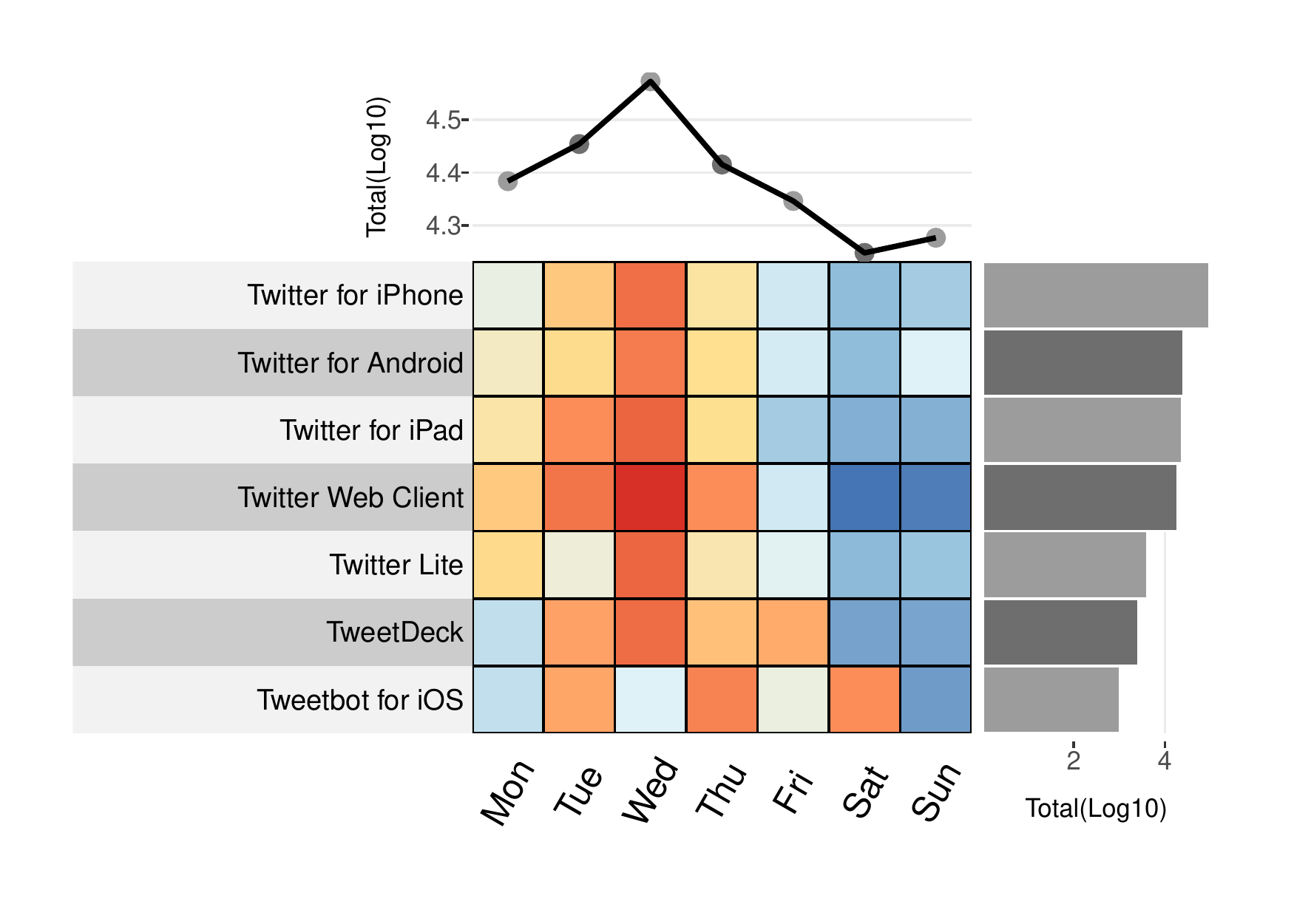}
		\caption{MPs' outgoing (\outgoing) activity by day of week  and Twitter client used. The `row bar' on right side and `col line' on top represents the counts of total data by Twitter client and day of week respectively. Dark blue represents the lowest activity and dark red the highest. Lowest activity is found trivially on Saturdays and Sundays and the highest activity is on Wednesdays, corresponding to Prime Minister's Questions. Android, iPhone and iPad clients are the most popular.}
		\label{fig:weekActive}
		\end{center}
		\vspace{-0.5cm}
	\end{figure}

The previous subsection identified selective responses and prioritisation of constituents as one way for MPs to cope with the load of engaging on Twitter. As an alternate or complementary strategy, MPs may also employ staff designated as Communications Officer or Senior Communications Officer. Permitted (non-party political) activities of such staff include establishing a social media presence in the constituency, publicising surgeries, following up on social media queries and comments, publicising the MP's parliamentary duties on social media and proactive and reactive communications with all media~\citep{MPsExpenses}. 

\begin{figure}
		\begin{center}
		\includegraphics[width=0.5\columnwidth]{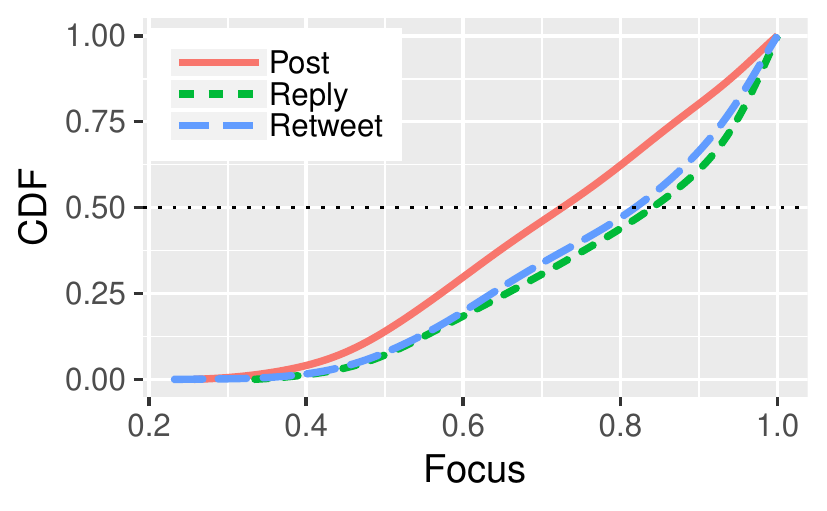}
		\caption{Cumulative Distribution Function (CDF) of the fraction of an MP's activity which is captured by their most commonly used source. For over half the MPs, more than 70\% of their original tweets (posts), and 80--90\% of their retweets and replies come from one source, which could imply  one person managing their handle.}
		\label{fig:sourceFocus}
		\vspace{-0.5cm}
		\end{center}
	\end{figure}

We cannot determine with certainty which tweeting instances originate from MPs and which from their staff, but we can find suggestive evidence. For instance, if multiple people are managing an account, it has the potential to be detected as a bot by the Botometer tool~\citep{davis2016botornot,varol2017online}. Only 45 of the 559  MPs ($\approx8\%$) in our data are detected as bots by this API\footnote{\url{https://osome.iuni.iu.edu/}}. We can also look at the Twitter client(s) used, as identified by the `statusSource' field of Tweets collected from the MPs' handles. A total of 47 different sources are used by the 559 MPs, ranging from the Twitter Web Client and TweetDeck, to Twitter for iPhone or Android. 
Fig.~\ref{fig:weekActive} shows that posting activity is mainly through the Web, whereas replies and retweets happen through personal devices such as iphone or android smartphones. Web clients can potentially come from multiple computers belonging to different staff. We also find that MP Twitter handles which use iPhone do not tend to also use android, and vice versa. Furthermore, the Web Clients are active mainly during weekdays. These patterns are suggestive of the MPs themselves, or one selected member of their staff handling the responses (replies and retweets), with the possibility of multiple staff being delegated the duty of posting new tweets, which may consist of advertising the MPs' activities; sharing videos and transcripts of their speeches etc. Note that the highest activity for the Web Client is on Wednesday, corresponding to Prime Minister's Questions, which, as mentioned before, is a highly advertised and popular activity. We also find that for over half the MPs, more than 80\% of their replies and retweets come from one source (Fig.~\ref{fig:sourceFocus}), which is further indicative of one person managing their Twitter presence.

These observations can potentially be explained by the rule that 
MPs may not claim for party political and campaigning activities \citep{MPsExpenses}.  
%
Some activities identified above, such as retweeting their party position, may not be allowable due to this rule, and would therefore need to be undertaken by the MP rather than their staff. This hypothesis is in alignment with our finding in Fig.~\ref{fig:sourceFocus}  that posts (original Tweets by MPs) are more likely to come from multiple sources than retweets -- recall that posts  tend to advertise MPs' parliamentary duties such as speeches and remarks made in the House of Commons, whereas  retweets tend to amplify messages of other party members or the official party handle.  MPs are also more likely to spend their limited budgets on communications staff only if certain conditions are met, for example MPs who are more junior and need to advertise themselves, or are in marginal seats~\citep{auel2018explaining}.
%



\section{Tone of political discussion}
Finally, we move to the third research question, and inquire about the nature and tone of the Twitter conversations. We are motivated to understand whether this platform, which appears to have gone mainstream in just five years since the first studies, and is  being widely used by nearly all MPs, is contributing positively to the political debate. 

This is an important question to answer, as various events such as Brexit have led to a highly charged and polarised political atmosphere in the UK, and both scholars and broadsheet newspapers have argued that the ``middle has fallen out'' of UK Politics~\citep{times,wpost,lse,opendemocracy}. There is also wide concern that political discussion on Twitter involves aggressive and ``trashy'' language \citep{guardian2016trashtalk,conover2011political}. 

Given the large scale of our data, we take a broad-brush approach, and focus on understanding whether there is cross-party political discussion between MPs and citizens, and on the tone and sentiments of the discussion as discoverable by tools such as LIWC 2015~\citep{pennebaker2015development}.

 \subsection{Cross-party political conversations}


\begin{figure}
\centering
\vspace{-0.0cm}
\includegraphics[width=0.5\columnwidth]{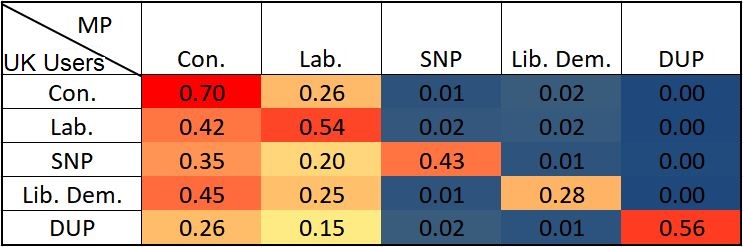}
	\caption{\textbf{Cross-party conversations in\ \incoming} Fraction of mentions from citizens who support (follow) MPs from one party to MPs of other  parties. Each row adds up to 1, including mentions from citizens to MPs of their own party.}
	\label{fig:UserCross}
	 \vspace{-0.5 cm}
\end{figure}

\begin{figure}
\centering
\includegraphics[width=0.5\columnwidth,height=0.18\columnwidth]{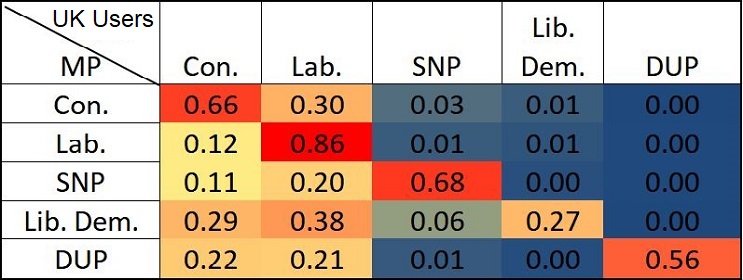}
	\caption{\textbf{Cross-party conversations in\ \outgoing} Fraction of MPs from one party replying to citizens who follow (support) MPs of other parties. Each row adds up to 1, including replies from MPs to citizens of their own party.}
	\label{fig:MPCross}
    \vspace{-0.5cm}
\end{figure}

To quantify polarisation, we divide users based on the party they support (using the method specified in the dataset section), and ask the extent to which they interact with MPs of other parties. Our focus on communications between people in power (MPs) and ordinary citizens distinguishes us from previous work that looked at how ordinary users have polarised~\citep{yardi2010dynamic,garimella2017long,borge2015content,conover2011political}.  Additionally, the UK is a multi-party system, which provides an interesting differentiating dimension to prior work, which has typically looked at a two-way polarisation, focusing on two sides of a conflict~\citep{yardi2010dynamic,borge2015content}, or on two-party systems like the USA~\citep{conover2011political,garimella2017long}.

 In Figure~\ref{fig:UserCross}, we examine Tweets from UK users that mention MPs, and find that regardless of the party they support, there is a lot of cross-party talk. Specifically, we focus on the top five parties in terms of MP numbers, and find that supporters of all parties tend to tweet mentioning MPs of the Conservative party, which is currently in power. Exploration with LDA topic modelling (not discussed in the paper) suggests that citizens are interested in topics such as the budget, Brexit, and resignations of ministers (due to scandals during the period of our collection). All of these have a natural focus on  ministers and MPs of the governing party, which helps explain the surprising amount of cross-party mentions. Conservative supporters have the largest proportion of within-party  mentions (69.8\%). This suggests that users' \textit{Tweets are directed at topical and current issues, and people involved in those issues, rather than the MPs they follow and the party they support, indicating a healthy  attitude of engagement beyond the ``echo chamber'' of people who have similar views} in online conversations . 
 
 An extreme example of cross-party conversation is the case of the Liberal Democrats (Lib Dems), whose supporters have more mentions of  Conservative MPs than  the 12 sitting MPs whose Twitter accounts they follow (Fig.~\ref{fig:UserCross}). In turn, Lib Dem MPs talk more to Labour supporters (who are more numerous) than those that follow them (Fig.~\ref{fig:MPCross}). A similar discrepancy between the interests of the MPs and its party supporters is observed with MPs of the Scottish National Party (SNP), whose replies have a greater proportion of replies to Labour supporters than conservative supporters, whereas ordinary citizens who follow SNP MPs talk more with conservative MPs than to Labour MPs. SNP and Lib Dems are ideologically closer to Labour than the Conservative Party\footnote{https://yougov.co.uk/news/2014/07/23/britains-changing-political-spectrum/, https://www.politicalcompass.org/uk2017}. We therefore conjecture that the discrepancy may be caused by MPs replying to those of a similar ideology as them, whereas citizens, who take a more questioning attitude (see next section), are engaging directly with the opposing view of the Conservatives.

 \begin{figure*}
	\centering
 	\includegraphics[width=1\textwidth]					{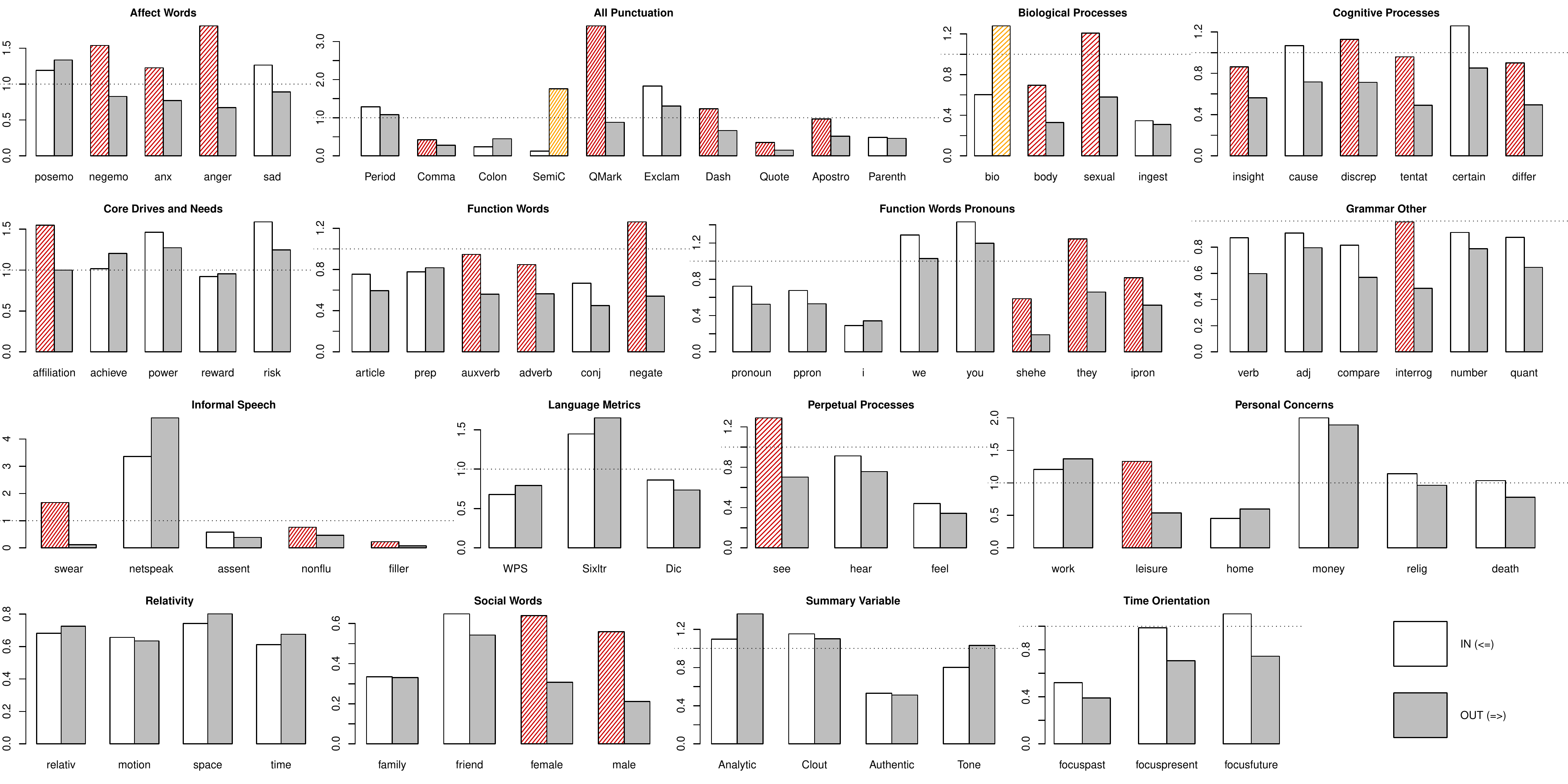}
	\caption{LIWC scores for variables or categories reported by LIWC. Scores are normalised by dividing by the base rate scores expected by LIWC \protect\footnote; if score is greater (resp.\ less) than 1, as marked by the horizontal dotted line, LIWC score is more (resp.\ less) than base rate. Corpus of \incoming\ tweets mentioning MP names is marked white; \outgoing\ tweets from MPs to citizens are marked gray. A category is marked in red (resp.\ orange) if  the score for\ \incoming\ (resp. \outgoing) is 50\% more than for\ \outgoing\ (resp.\ \incoming) corpus.}
	\label{fig:liwc}
\end{figure*}

\footnotetext{\url{http://liwc.wpengine.com/compare-dictionaries/}}

 \subsection{Language and sentiments}
 
Given the surprising result of substantial cross-party talk between MPs and users, it is natural to ask what the tone of the conversation is between MPs and citizens -- i.e., whether the discussion between MPs and citizens is a civilised discussion or an aggressive slanging match as in recent political campaigns~\citep{guardian2016trashtalk}.  To measure this, we use LIWC to summarise the language of the citizens mentioning MPs (\incoming) on the one hand, and MPs replies to citizens' Tweets (\outgoing) on the other. We look for differences and similarities in language usage between these two categories, to understand the tone of the discourse between MPs and citizens.

LIWC provides 94 dimensions along which to measure different aspects of language use~\citep{pennebaker2015development}. For each dimension, it also gives the base rates of word counts to be expected in normal usage. Fig.~\ref{fig:liwc} shows the LIWC scores obtained for each dimension of language use, for both ~\incoming\ and ~\outgoing, normalised by the expected base rates of usage of that dimension. 

To take a principled approach, we focus on the LIWC categories which see higher than base rates of usage or where there is substantial ($>50\%$) difference between ~\incoming\ and ~\outgoing. Using this approach, we can make the following observations from Fig.~\ref{fig:liwc}: 
\begin{enumerate}
\item Both MPs and citizens show more positive emotions than LIWC base rates, which could be suggestive of a respectful or appreciative discussion. 
\item However, citizens' incoming (\incoming) Tweets  also express more than the base rate of negative emotion, anxiety and anger, suggesting more conflict in some conversations. The citizens' Tweets also raise a lot of questions, heavily using interrogatives, and question marks in their language.  This could potentially be related to  scandals and related resignations during the period of study.
\item Unusually for political conversations, there is a large amount of sexual and sex-related words, owing to the Westminster sex scandal which erupted in the wake of the \#MeToo movement, and led to the resignation of several ministers and MPs during the period of our study\footnote{Wikipedia provides the most up to date account at \url{https://en.wikipedia.org/wiki/2017_Westminster_sexual_scandals}. Sky News also has a useful timeline of one of the most intense weeks: \url{https://news.sky.com/story/a-week-in-westminster-sex-scandal-timeline-11111086}}. 
\item Perhaps because of this scandal, citizens' incoming (\incoming) tweets towards MPs have a higher than base rate of ``moralising'' language, using words such as `should', `would' (marked as discrep), and `always', `never' (marked as certain).
\item Words relating to power and risk figure highly in the MPs' language as well as the citizens'.
\end{enumerate}

 \begin{table}
 \centering
 	\begin{tabular}											{|p{1.5cm}|p{1.5cm}|p{1.5cm}|p{1.5cm}|} 
		\hline
		Dimension & Difference &
        Dimension & Difference \\
		\hline
        Swear &90.5\%&Nonflu&63.2\%\\
        QMark &79.2\%&Shehe&63.2\%\\
        Filler&78.4\%&Anger&58.2\%\\
        Exclam&72.2\%&Leisure&58.1\%\\
        Negate&66.4\%&Male&56.6\%\\
        \hline
	\end{tabular}
 \caption{LIWC categories which make it less likely that MPs respond back to citizen. This table shows the percentage difference of some LIWC dimensions that corresponds to the decreased likelihood of MPs in making responses to incoming ($\Leftarrow$) mentions if these LIWC categories are present.}
 \label{tbl:what-elicits-response}
 \vspace{-0.5 cm}
 \end{table}

Table~\ref{tbl:what-elicits-response} shows that MPs appear to take into consideration the language of a mention in deciding whether to reply back. Incoming (\incoming) mentions from citizens which exhibit anger or use swear words, question marks, filler words and other non-fluencies (e.g., ``err'', ``um'', I mean, you know, etc...) are less likely to elicit a reply.
 
Our broad-brush approach is intended only to provide a flavour of the tone of discussion. It appears to indicate that in our study period, which was rich in scandals that affected multiple MPs and ministers, citizens are using Twitter as a platform to freely and directly question their representatives and express  negative emotions, anxiety and anger, as they are entitled to. However, they also show higher than base rates of positive affect and appreciation where warranted. In return, MPs appear to exercise restraint, using higher than base rates of positive language, and avoiding using or responding to negative language (which could escalate conflict). We conjecture that the public nature of Twitter leads to MPs being conscious of the effect of their words on their image and public perception, providing a platform for civilised discourse. We note that this kind of behaviour may be partly due to our focus on interactions between MPs and citizens. Previous studies in the UK context have found that MPs have indulged in attacks on other politicians, especially in election contexts~\citep{graham2013between}.

\section{A possible future of online Twitter engagement}

We conclude with a brief case study of a novel way in which the immediacy of Twitter was used to improve democracy by allowing citizens a part in creating an Act of Parliament, and discuss how it speaks to our three research questions: 
%

Individual MPs in the UK Parliament are able to submit Bills (also known as draft legislation).  These are known as Private Members' Bills.  Priority is given to Government-sponsored Bills, so to ensure that a proportion of Private Members' Bills have a chance to become law, there is a ballot of MPs each year to assign priority for the limited amount of debating time available.  However, even Bills coming high up in the ballot are unlikely to be passed unless they have the tacit or explicit support of the Government.
 
 
Chris Bryant, a Labour back bench MP, came top in the ballot for the 2017-19 session, and therefore was eligible to propose a Bill. However, as a member of the Opposition Party who is also not among the prominent ``front bench'' MPs, his bill would have faced an uphill task. Bryant launched a consultation on Twitter in July 2017 (before our study period), putting forward six possible Bills and asking Twitter users to choose their favourite through an online survey. 45,000 people participated\footnote{https://www.youtube.com/watch?v=RqkSSQ3bAaA}, and the winner of the poll was a proposal to provide additional legal protection to emergency service workers, as a result of reports of assaults by members of the public during emergency call-outs.   Bryant introduced the Assaults on Emergency Workers (Offences) Bill 2017-19. This bill was greatly strengthened by the evidence of public support, and was one of the few Private Members' Bills supported by the Government. Tweets during the passage of the Bill through Parliament used the hashtag \#ProtectTheProtectors, and received a high level of engagement.   On 13 September 2018, the Bill received Royal Assent, the final stage on the way to becoming a law. It is now been signed into law as the Assaults on Emergency Workers (Offences) Act 2018.

\begin{figure}
	\begin{center}		\includegraphics[width=.4\columnwidth]{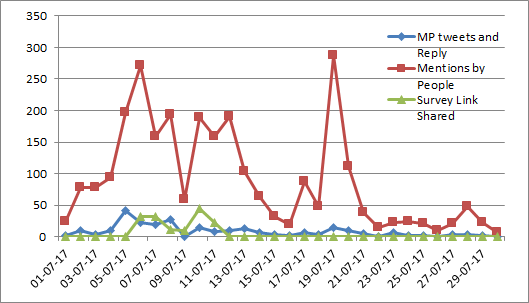}
		\includegraphics[width=.25\columnwidth]{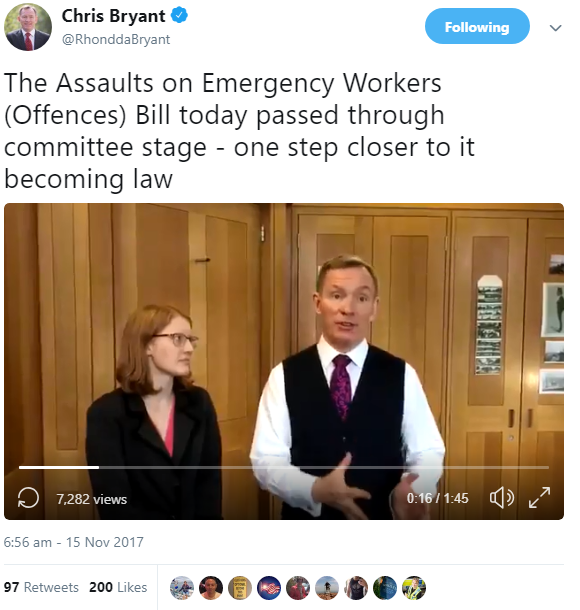}
		\caption{Left: Survey tweet links, MP Chris Bryant $\Rightarrow$ and $\Leftarrow$ Right: MP Chris Bryant's  thanking video after his proposed bill was passed the Committee stage.}
		\label{fig:billSurvey}
	\end{center}
\vspace{-0.5 cm}
\end{figure}

This innovative approach, and the effective use of Twitter, surveys and hashtags has enabled the public to follow the progress of the Bill throughout its timeline and  participate by providing direct comments, creating an experience closer to direct democracy~\citep{coleman2005blogs}. It has also acted as a means of garnering publicity for Bryant. We can relate this case back to the three research questions: RQ-1 and RQ-2 seek to understand how much load is incurred by the MPs, and how they manage this load. Clearly, with 45,000 responses to the initial survey, this was a huge effort. The use of a survey tool was critical to manage this huge load and summarise their response. However, the MP also struggled to cope: Analysis of Tweets during the survey/poll in July 2017  reveals a typical pattern of activity during a focus period (Fig.~\ref{fig:billSurvey}).  Twitter users mentioned the MP (in red), often to make suggestions or enter into dialogue, but as shown by the blue line, he was unable to respond to many.  This showcases both the potential for direct and participatory online engagement via Twitter but also the drawbacks, if excessive activity makes a personal response impracticable.  

\begin{figure}
		\begin{center}		\includegraphics[width=.4\columnwidth]{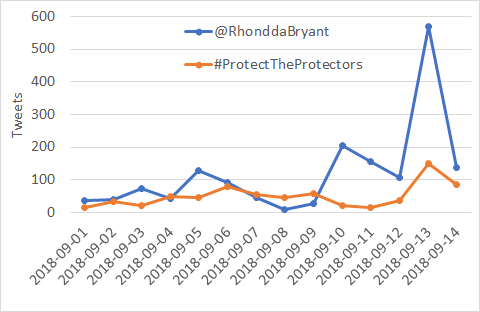}
\includegraphics[width=.35\columnwidth]{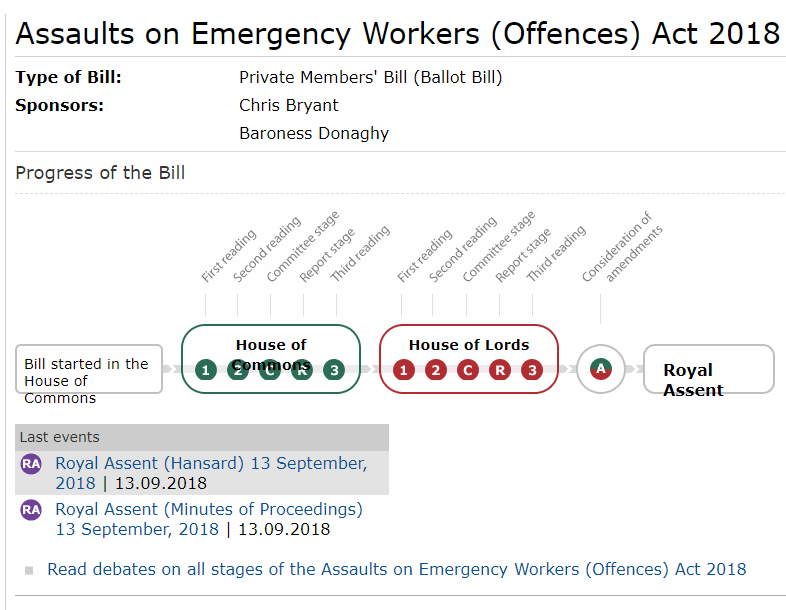}
		\caption{Left:\#ProtectTheProtector tweets burst on the day of Bill getting passed in House of Commons. Right: Status of the Bill\protect\footnote as on 15$^{th}$ Sep 2018.}
		\label{fig:bill}
		\end{center}
	\end{figure}

To study RQ-3 on the nature and tone of the discussion, we focus on the final day of high activity during the passage of the Bill. Fig.~\ref{fig:bill} shows that the MP gained more than 500 mentions on just one day (Sep 13), when the Royal Assent was obtained. With a high number of mentions like this on a single day, it is hard to respond  to each mention; reiterating again that although the process innovatively unlocked the  participatory potential of Twitter, the burden of response during such direct engagement remains an issue (RQ1). To manage the load, the MP did a `thank you' post as a collective response to all (RQ2). As an event where high attention was anticipated, the messages were mostly appreciative and complementary to the MP\footnote{The volume of the whole conversation from July 2017--Sep 2018 permits only a cursory examination, but also seems to incorporate a mostly civil and respectful tone; with suggestions and requests for changes, inquiries as to why the bill is required when assault by itself is already considered a crime, as well as  messages of encouragement.}. Although as expected the majority of congratulatory Tweets were from Labour supporters, it is remarkable that close to 28\% of tweets come from non-Labour supporters, showing the broad multi-partisan support for the Bill, offering hope for constructive participatory democracy through the innovative use of digital tools like Twitter.

\footnotetext{\url{https://services.parliament.uk/bills/2017-19/assaultsonemergencyworkersoffences.html}}

\section{Discussion and further work} \label{sec:conclusion}

As early as 1774, the political philosopher and MP Edmund Burke stressed the importance of understanding the views of constituents~\citep{Burke}. 
Despite this early recognition of its need, active engagement with constituents outside the election period was rare until the mid-twentieth century~\citep{auel2018explaining}. However, it is now seen as a necessity by MPs, and this is being increasingly facilitated through digital means. 

Our research agenda involves understanding the usage of Twitter as a new form of continuous citizen engagement. In this work, we identified Twitter as an interactive platform which seems to have become part of mainstream usage,  used by nearly all MPs, and with a high volume of activity. We investigated the dynamics of the load imposed by the increasing volumes of Twitter activity and the consequent attention directed towards MPs. We showed that attention can be highly focussed, with a large proportion of the total activity directed at an MP occurring during short focus periods of 3--5 days. MPs use selective replies and prioritisation of local or constituents' concerns as a way to manage this high attention load. They use their Twitter presence strategically, balancing their role as party representatives with the role of hearing and responding to their citizens' needs. We also find that Twitter presents possibilities for immediate and direct discussion,  leading to new possibilities for cross-party discussions on a level playing field and therefore holds promise for bridging, or at least initiating conversations, across the political divide.

However, there is additional work to be done. For instance, despite its current widespread use, there still remain some concerns about how representative Twitter is, as a (or the main) platform for digital citizen engagement. Furthermore, we need to go beyond the current observational study, to conclusively understand whether Twitter engagement is helping MPs in their day-to-day duties or if it merely adds to their burdens. Other questions -- such as whether Twitter remains a place for ``empty'' conversations, or whether actual Government or Parliamentary activity result from these online discussions -- need closer scrutiny. Case studies such as the creation of the Assaults on Emergency Workers (Offences) Act 2018 point to ways in which Parliamentary activity can be facilitated or directed through Twitter and other online means, but these are early examples, and there may be other mechanisms that become more commonplace in the near future. In this context, it may be interesting to compare with other more formal routes, including e-petitions, which can get discussed in parliament if sufficient numbers of citizens declare interest in an issue.

%
%
%
%
%
{
	\balance{
		\bibliographystyle{unsrt}
		\bibliography{TweetingMPs_Pushkal_Arxiv} 

	}
}


\end{document}